\documentclass[10pt]{article}

\usepackage{graphicx}
\usepackage{dcolumn}
\usepackage{bm}
\newcommand{\be}{\begin{equation}}
\newcommand{\ee}{\end{equation}}
\newcommand{\bea}{\begin{eqnarray}}
\newcommand{\eea}{\end{eqnarray}}
\newcommand{\nn}{\nonumber}

\begin{document}
\begin{center}
{{\LARGE \bf Radiative Processes in Quark-Gluon Plasma}} \\
\bigskip
{\large \bf Trambak Bhattacharyya\footnote{email: trambakb@vecc.gov.in}, Surasree Mazumder and Jan-e Alam} \\
{Variable Energy Cyclotron Centre, Kolkata, India} 
\bigskip
\end{center}
\begin{abstract}
The spectrum of emitted gluons from the process $\mathrm{gg\rightarrow ggg}$ 
has been evaluated by relaxing some of the approximations used in earlier works.
The formula obtained in the present work has been applied to several physical 
quantities. A general expression for the dead cone of gluons radiated 
by virtual partons has been derived. It is observed that the suppression caused by 
the high virtuality is overwhelmingly large as compared to that on account of 
conventional dead-cone of heavy quarks.
\end{abstract}
\section{Introduction}
Radiative processes like $\mathrm{gg\rightarrow ggg}$ are of particular importance
in the study of quark gluon plasma (QGP)expected to be formed in heavy ion collisions (HIC)
at ultra-relativistic energies.  The process,
 $\mathrm{g+g\rightarrow g+g+g}$ has drawn 
particular attention in view of its importance for the 
chemical equilibration,
 energy loss of gluons in QGP, evaluation of transport 
coefficients of the gluonic plasma etc. There
has been recent attempts ~\cite{DA,GR} to generalize the 
Gunion-Bertsch (GB) formula ~\cite{GB} for gluon rediation from light partons. 
Our aim is to find correction terms to GB formula relaxing earlier
approximations and to see the effects of the correction terms on (i) equilibration rate 
and (ii) the energy loss of gluons in gluonic plasma. Also, it is expected that the 
radiated soft gluon spectrum emitted from heavy quarks will be supressed compared to that
emitted from light quarks~\cite{khar}. A new suppression mechanism due to high virtuality of quarks 
has been proposed and it is shown that gluon emitted from virtual quarks, irrespective of their masses, 
are exposed to this suppression. The dead cone due to virtuality
may play a crucial role in explaining the observed similar suppression patterns 
of light and heavy quarks jets in heavy ion collisions at Relativistic Heavy Ion Collider (RHIC)~\cite{stare}.
\begin{figure}[h]                                                            
\begin{center}                                                               
\includegraphics[scale=0.40]{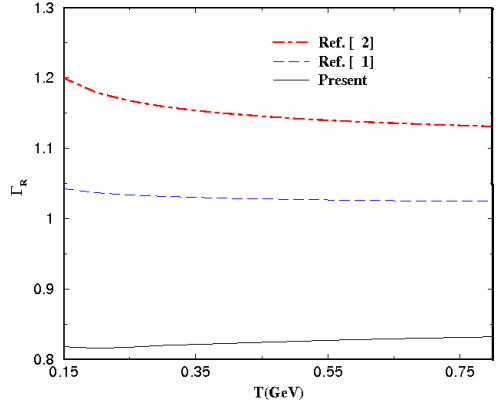}                                 
\caption{(Color online) 
Temperature variation of the ratio of the equilibration rate  
(inverse of the time scale) 
obtained in the present work (solid line), Ref.~\cite{DA} (dashed line),
and  ~\cite{GR} (dot-dashed) normalized 
by the GB value 
for the process $\mathrm{gg}\rightarrow \mathrm{ggg}$. 
}
\label {fig1} 
\end{center}                                                               
\end{figure}                                                                
\begin{figure}[h]                                                            
\begin{center}                                                               
\includegraphics[scale=0.40]{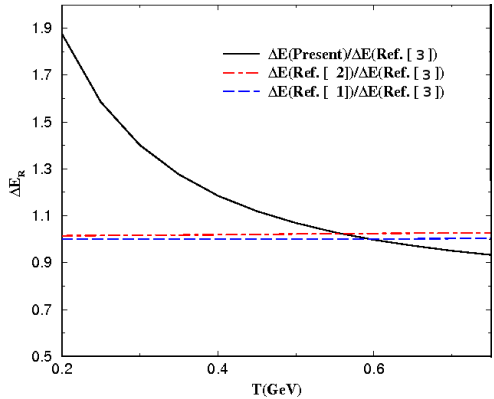}                                 
\caption{(Color online) Temperature variation of $\Delta E_R$ of a 15 GeV gluon 
moving through a gluonic heat bath of dimension 4 fm. Solid (dashed) line indicates result for
the gluon spectrum obtained in the present work (~\cite{DA}). 
The dot-dashed line stands for the results 
for the gluon spectrum of ~\cite{GR}.}                                                          
\label {fig2}  
\end{center}                                                                 
\end{figure}                                                                 
\begin{figure}[h]      
\includegraphics[scale=0.4]{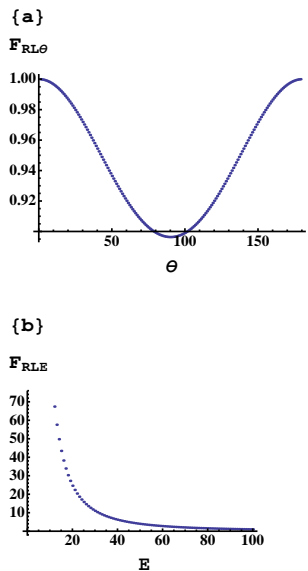}                                 
\caption{(Colour online) The variation of (a) $F_{\mathrm RL\theta}$ with $\theta$ and
(b) $F_{\mathrm RLE}$ with $E$ for virtual light quarks. 
}
\label{fig3}                                                               
\end{figure}                                                                 
\begin{figure}[h]      
\includegraphics[scale=0.4]{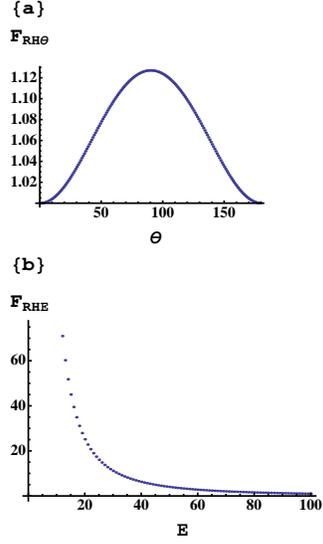}                                 
\caption{(Colour online) The variation of (a) $F_{\mathrm RH\theta}$ with $\theta$ and
(b) $F_{\mathrm RHE}$ with $E$ for virtual heavy quarks. }
\label{fig4}                                                               
\end{figure}                                                                 

\section{Correction to GB Spectrum}
Following ~\cite{berendes}, the matrix element for $gg\rightarrow ggg$, $|M_{\mathrm{gg\rightarrow ggg}}|^2$
after simplification can be written as~\cite{bmad}:
\bea
{|M|^2}_{\mathrm{gg\rightarrow ggg}}&=& 12g^2 |{M _{\mathrm{gg\rightarrow gg}}}|^2 \frac{1}{k_{\bot}^2} 
\times[(1+\frac{t}{2s}+\frac{5t^2}{2s^2}-\frac{t^3}{s^3})\nonumber\\
&&-(\frac{3}{2\sqrt{s}}+\frac{4t}{s\sqrt{s}}-\frac{3t^2}{2s^2\sqrt{s}})k_{\bot} \nonumber\\
&&+(\frac{5}{2s}+\frac{t}{2s^2}+\frac{5t^2}{s^3})k_{\bot}^2],
\eea
where
$|M_{\mathrm{gg\rightarrow gg}}|^2=(9/2)g^4s^2/t^2$,
$s=(k_1+k_2)^2, t=(k_1-k_3)^2$, $u=(k_1-k_4)^2$,
$k_{\perp}$ is the transverse momentum of the radiated gluon.
$g=\sqrt{4\pi\alpha_s}$ is the color charge, and
$\alpha_s$ is the strong coupling. Effect of the modified gluon distribution 
on energy loss of gluons in gluonic plasma and on its equilibration rate has 
been discussed in later sections.

\section{Dead-cone due to virtuality}
For the demonstration of suppression of soft gluon radiation due 
to virtuality, we take up the $e^+e^-\rightarrow Q\bar{Q}g$ process, 
where $Q$ is quark. 
The spectrum of the soft gluons emitted by the virtual quarks can be 
shown to be~\cite{ba}:
\bea
F&=&\omega^2(2R_{43}-R_{44}-R_{33})\nn\\
&&=4\beta^2 \left(\frac{\frac{V^4}{\omega^2 E^2}+\frac{4V^2}{\omega E}+4sin^2 \theta}
{(\frac{V^4}{\omega^2 E^2}+\frac{4V^2}{\omega E}+4(1-\beta^2 cos^2 \theta))^2}\right)~,
\eea
where we assume that external quarks are on the verge of 
being on-shell so that Dirac’s equation can be applied. $V$ is the `virtuality parameter'
defined by the equation, $V^2=q^2-m_Q^2$  
where $q^2$ is four-momentum square of external virtual particles, $q^2=m_Q^2$ implies $V=0$, i.e.
the particle becomes on-shell. We can show that the spectrum is that of gluons emitted from
on-shell quarks when $V=0$. $\omega$ is the energy of the soft gluon emitted at angle $\theta$
with the parent quark whose velocity is $\beta$ and enegry is $E$.
We replace the virtuality by $V=\sqrt{q^2-m^2}=\sqrt{E^2-p^2-m^2}$, where $p=\beta E$ and define
$F_{\mathrm RH\theta}=F(E=1.5 GeV,\theta)/F(E=1.5 GeV,\theta=0)$  for heavy quarks. A similar
quantity $F_{\mathrm RL\theta}=F(E=3 GeV,\theta)/F(E=3 GeV,\theta=0)$ is defined for light 
quarks. Similarly $F_{\mathrm RHE}=F(E,\theta=\pi/4)/F(E=100 GeV,\theta=\pi/4)$ and
$F_{\mathrm RLE}=F(E,\theta=\pi/4)/F(E=100 GeV,\theta=\pi/4)$ are two quantities just 
by choosing proper $F$(spectrum) for heavy or light quarks.

\section{Results and Discussion}
The relative energy loss ($\Delta E_R$), defined by energy loss of gluons ($\Delta E$) 
normalized by the corresponding value obtained from GB approximation
and relative equilibration rate($\Gamma_R$) 
of fast gluons, defined in the same way, are calculated and compared with ~\cite{DA,GR}. 
We observe that with the correction terms the value of $\Delta E_R$ is
enhanced by about $40\%$ and $20\%$ for $T$ 300 MeV and 400 MeV, respectively,
compared to the $\Delta E_R$ obtained from the spectra of  Refs.~\cite{DA} and ~\cite{GR}.
Such differences may have important consequences on the
heavy-ion phenomenology at RHIC and LHC collision energies.

Soft gluon spectrum ($\omega=30$ MeV) radiated by virtual light quarks,  
$F_{\mathrm RL\theta}$  with low virtuality($E=3$ GeV) varies differently (see Fig.~\ref{fig3})
with $\theta$ from the corresponding quantity, $F_{\mathrm RH\theta}$ for virtual heavy quarks (see Fig.~\ref{fig4}). 
This is obvious because for low virtuality the light 
partons are not subjected to any dead cone suppression at $\theta=0$ and $\pi$ 
unlike heavy quarks.
But the spectrum is similarly
suppressed with energy for heavy or light quarks. Since energy is measure of virtuality in our
formalism, we say that highly virtual quarks, heavy or light are always exposed to similar 
suppression. 

\section{Summary}
In summary, we have calculated the correction to Gunion-Bertsch spectrum relaxing 
earlier approximations and calculated energy loss as well
as equilibration rate of fast gluons passing through gluonic plasma
. We proposed a new radiative soft gluon suppression due 
to virtuality of parent quark. For high virtuality of quarks, gluons emitted from
them are exposed to similar radiative suppression.

\section{Acknowledgement}
Fruitful discussions with B. Z. Kopeliovich and S. Sarkar are 
acknowledged. TB and SM are supported by DAE, Govt. of India.

\end{document}